\documentclass[aps,prl,twocolumn,superscriptaddress,a4paper]{revtex4}

\usepackage{amsmath,amssymb,graphicx,dcolumn,bm,url}
\usepackage[colorlinks=true,linkcolor=red,citecolor=blue]{hyperref}

\begin{document}
\title{Quantum state estimation with unknown measurements}
\author{Merlin Cooper}
\email{m.cooper1@physics.ox.ac.uk}
\affiliation{Clarendon Laboratory, University of Oxford, Parks Road, Oxford, OX1 3PU, UK}
\author{Micha{\l} Karpi{\'n}ski}
\affiliation{Clarendon Laboratory, University of Oxford, Parks Road, Oxford, OX1 3PU, UK}
\author{Brian J. Smith}
\affiliation{Clarendon Laboratory, University of Oxford, Parks Road, Oxford, OX1 3PU, UK}
\maketitle
\textbf{
Improved measurement techniques are central to technological development and foundational scientific exploration. Quantum optics relies upon detectors sensitive to non-classical features of light, enabling precise tests of physical laws \cite{Zavatta:09, Yao:10, BachorRalphBook, LIGO:13} and quantum-enhanced technologies such as precision measurement \cite{LIGO:13} and secure communications \cite{O'Brien2009}. Accurate detector response calibration for quantum-scale inputs is key to future research and development in these cognate areas. To address this requirement quantum detector tomography (QDT) has been recently introduced \cite{DAriano:04,Coldenstrodt-Ronge:09,Lundeen:09,Amri:11,Zhang:12,Brida_njp:12,Natarajan:13}. However, the QDT approach becomes increasingly challenging as the complexity of the detector response and input space grows. Here we present the first experimental implementation of a versatile alternative characterization technique to address many-outcome quantum detectors by limiting the input calibration region \cite{Rehacek:10,Mogilevtsev:13}. To demonstrate the applicability of this approach the calibrated detector is subsequently used to estimate non-classical photon number states.
}

Experimental quantum optics relies upon the ability to create, manipulate, and measure the state of the light field for applications ranging from fundamental scientific research to the development of quantum technologies that harness the non-classical behaviour of light \cite{LIGO:13, BB84, Becerra:13, Taylor:13}. Methods to characterize each step of a quantum experiment are crucial to ensure appropriate evaluation of tests of theoretical predictions and desired device operation. Well-developed techniques for quantum state estimation (QSE) \cite{LeonhardtBook, Paris:04, Lvovsky:09} and quantum process tomography (QPT) \cite{OBrien:04,Lobino:08} rely upon accurate knowledge of the detector response. Only recently has the independent characterization of quantum detectors with few outcomes been experimentally demonstrated \cite{Lundeen:09, Zhang:12,Brida_njp:12, Natarajan:13} by means of quantum detector tomography (QDT). 

Complete characterization of the detector response through QDT becomes increasingly demanding as the number of detector outcomes grows. The primary roadblocks arise from the need to acquire and analyse expanded data sets, which becomes intractable with current experimental and computational capacity. This is due to experimental instability over the time required to collect sufficient data, the size of the numerical inversion problem and unavoidable statistical noise that can distort rare detection events \cite{Zhang_NJP:12}. Here we present the first experimental demonstration of an alternative approach, known as the fitting of data patterns (FDP) \cite{Rehacek:10,Mogilevtsev:13}, that enables calibration of detectors with a sizable number of outcomes and their subsequent use in state estimation. The FDP method is applied to a balanced homodyne detector (BHD) \cite{LeonhardtBook, Lvovsky:09}, a central resource in a broad class of quantum optical experiments \cite{Braunstein:05}, that has yet to be independently characterized. The BHD employed here has more than 150 outcomes, which is an order of magnitude more than any quantum detector characterized to date. To demonstrate the FDP method as a tool for complex detector calibration and quantum state estimation, we subsequently present QSE of non-classical states of light using the independently calibrated BHD. 

\begin{figure*}[t]
\centerline{\includegraphics[width=1.00\textwidth]{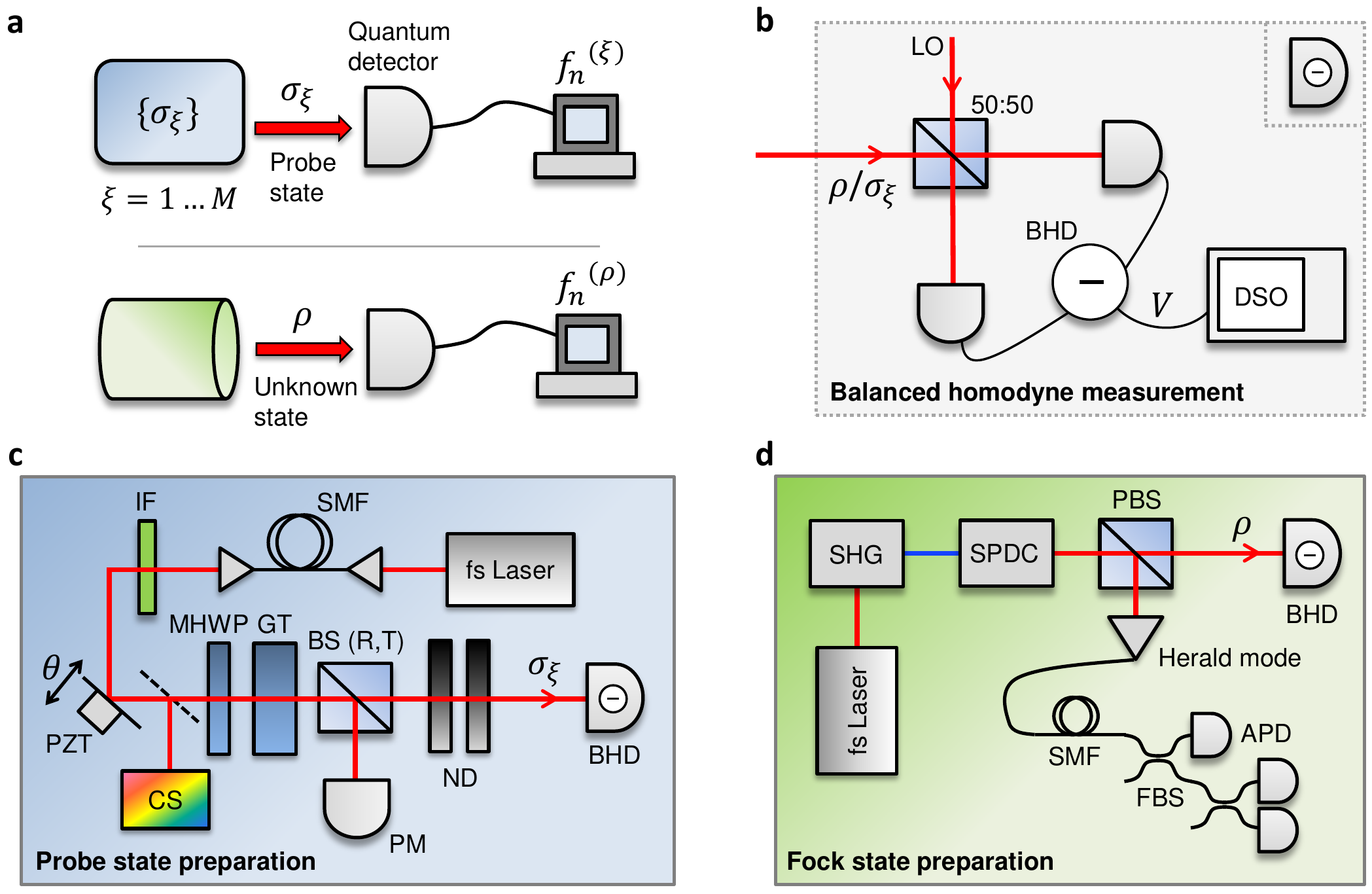}}
\caption{\textbf{a} Conceptual and \textbf{b-d} experimental diagrams of the FDP procedure. \textbf{a} top: multiple copies of known probe quantum states $\{\sigma_{\xi}\}$ are separately measured by the unknown detector to form the data patterns $\{f_{n}^{(\xi)}\}$. Bottom: multiple copies of the unknown quantum state $\rho$ to be estimated are measured giving the frequency distribution $f_n^{(\rho)}$. \textbf{b} Balanced homodyne detection: probe states $\{\sigma_{\xi}\}$ or unknown state $\rho$ are combined with the local oscillator (LO) on a 50:50 beam splitter. The output modes are directed to a pair of photodiodes whose photocurrents are subtracted, with the resulting difference current converted to a voltage measured with a digitizing oscilloscope (DSO) \cite{Cooper_jmo:13}. This yields a set of voltages $\{V\}$ for each state from which the frequency distributions $\{f_n^{(\xi)}\}$ and $f_n^{(\rho)}$ are determined. \textbf{c} Probe state preparation: The output of a pulsed Ti:Sapphire laser oscillator is spatially and spectrally filtered with a single-mode fiber (SMF) and interference filter (IF). A motorized half-wave plate (MHWP) and Glan-Taylor polarizer (GT) followed by neutral density (ND) filters control the probe state amplitude, while a piezoelectric transducer (PZT) averages the phase. A calibrated power meter (PM) and compact spectrometer (CS) monitor the optical power and wavelength, respectively, to determine the probe state amplitude. \textbf{d} Fock state preparation: The Ti:Sapphire laser output is frequency doubled (SHG) to pump a spontaneous parametric down-conversion source (SPDC). A spatially multiplexed detector comprising a fiber beam splitter (FBS) network and three avalanche photodiodes (APDs) enables heralding of multi-photon Fock states $\rho$.}
\label{fig:schematic}
\end{figure*}

\begin{figure*}[t]
\centerline{\includegraphics[width=1.00\textwidth]{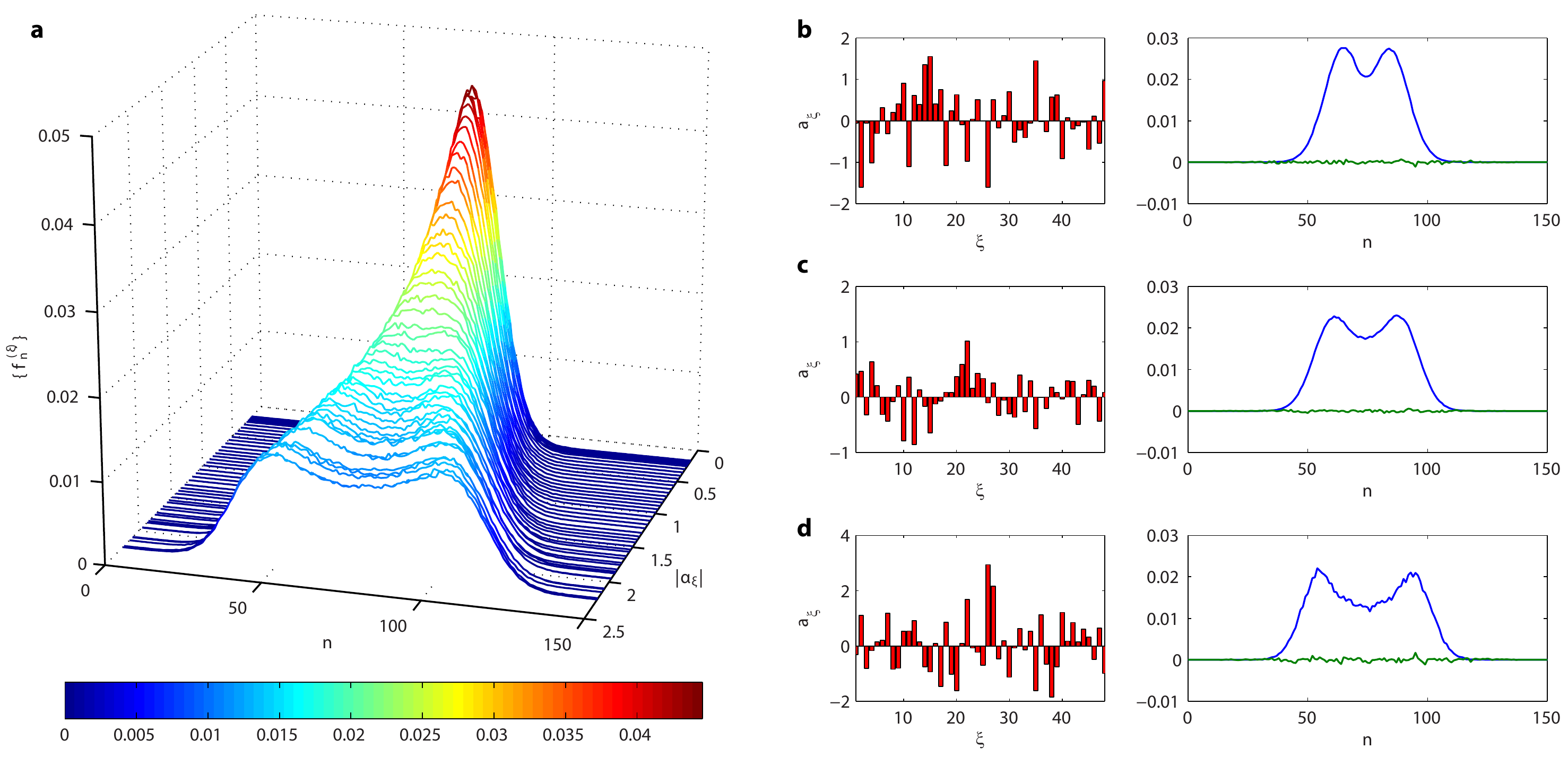}}
\caption{\textbf{a} Data patterns $\{f_{n}^{(\xi)}\}$ obtained for the set of 48 PHAV probes $\{\sigma_{\xi}\}$ with $|\alpha_{\xi}|$ ranging from 0.17 to 2.24. \textbf{b-d} left, optimal coefficients $\{a_{\xi}\}$ minimizing the functional defined in Eq.~(\ref{eq:functional}) for the one-, two- and three-photon Fock states. \textbf{b-d} right, measured frequency distributions $f_{n}^{(\rho)}$ (blue curves) and residuals between fitted frequency distributions and measured frequency distributions $f_n^{(\rho)}-\sum_{\xi} a_{\xi} f_{n}^{(\xi)}$ (green curves). The latter give a measure of the deviation between the measured and predicted frequency distributions, which is found to be within the statistical noise due to the finite number of measurements. }
\label{fig:probes}
\end{figure*}
\begin{figure*}
\includegraphics[width=1.00\linewidth]{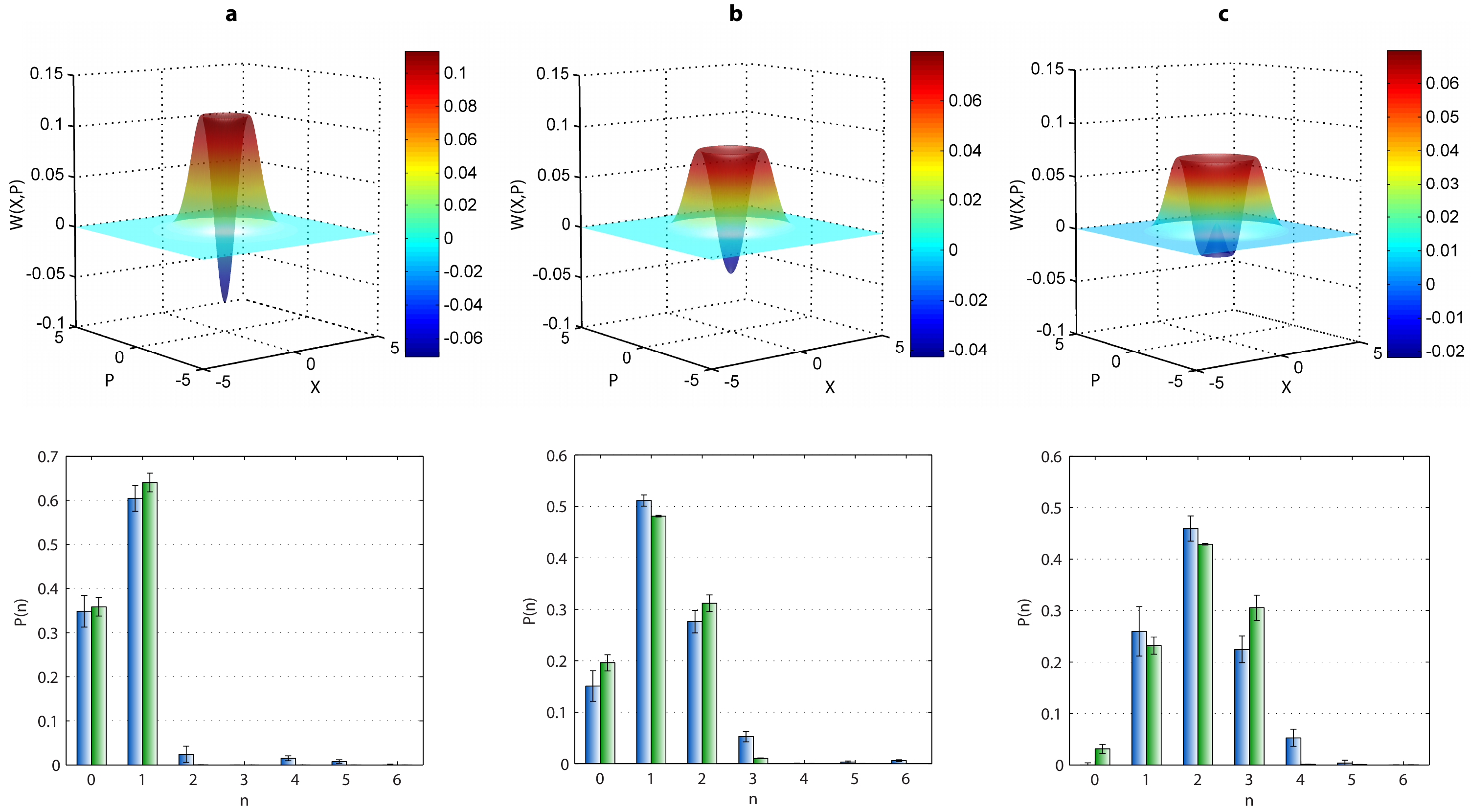}
\caption{Reconstructed Wigner functions $W(X,P)$ from FDP method (top) and photon number statistics $P(n)$ (bottom) from FDP (blue bars) and ML (green bars) methods for \textbf{a} $n=1$, \textbf{b} $n=2$ and \textbf{c} $n=3$ Fock states. The reconstructed Wigner functions exhibit negative values, thus indicating the non-classical nature of the reconstructed states. Error bars indicate one-$\sigma$ confidence intervals and include both systematic and statistical sources of error for both the FDP and ML state reconstructions (see Methods, ``Reconstruction and uncertainty estimation'').}
\label{fig:density_matrices}
\end{figure*}

In quantum theory the detector response is mathematically represented by its positive operator value measure (POVM) comprising a set of positive operators $\{\pi_n\}$, with $n=1,2,...N$, labelling the measurement outcomes. The probability for detector outcome $n$ given input state $\sigma_{\xi}$ is determined by the Born rule 
\begin{align}
p_n^{(\xi)}=\text{Tr}(\pi_n \sigma_{\xi}).
\label{eq:bornrule}
\end{align}
Quantum detector tomography (QDT) aims to determine the set of measurement operators $\{\pi_n\}$ by experimentally estimating the outcome probabilities $\{p_n^{(\xi)}\}$ for a set of known probe states $\{\sigma_{\xi}\}$, and inverting Eq.~(\ref{eq:bornrule}) \cite{DAriano:04,Lundeen:09,Amri:11,Coldenstrodt-Ronge:09,Zhang:12,Brida_njp:12,Natarajan:13}. However, as the number of measurement outcomes and input state space increase, so do the experimental and computational tasks associated with QDT \cite{Zhang:12,Zhang_NJP:12}. The FDP technique limits the input state space to a region of interest set by the experimenter, enabling a substantial reduction in the experimental and computational challenges. Furthermore, unlike QDT which requires significant numerical post-processing involving optimization techniques or model fitting, the data patterns are obtained directly from the experimental detection events. 

The FDP approach begins by characterizing the detector response to a set of $M$ known input probe states $\{\sigma_{\xi}\}$ that span the input region of interest. Here $\sigma_{\xi}$ is the density operator for the state labelled by $\xi=1,2...,M$. Multiple copies of each probe state are sent to the detector and the frequency distribution $f_n^{(\xi)}$ of measurement outcomes, labelled by $n$, associated with input state $\sigma_{\xi}$, is determined as depicted in Fig.~\ref{fig:schematic}(a). This procedure is similar to mapping the impulse response function of a linear optical system. The data pattern set $\{f_n^{(\xi)}\}$ constitutes the detector calibration within the field of view defined by the probe states. As with QDT, the FDP method relies upon the ability to reliably produce a sufficient set of well-known probe states to characterize the detector response \cite{Rehacek:10,Mogilevtsev:13}.

To estimate an unknown quantum state with density operator $\rho$ using the FDP approach, many identically prepared copies of the state are measured with the calibrated detector. The frequency distribution of measurement outcomes for the unknown state $f_n^{(\rho)}$ is determined from the acquired data. The unknown state density operator is estimated as a weighted sum of the probe states
\begin{align}
\rho^{\rm{fdp}} = \sum_{\xi} a_{\xi} \sigma_{\xi},
\label{eq:estimatedstate}
\end{align}
where the real-valued expansion coefficients $\{a_{\xi}\}$ are found by minimizing the functional 
\begin{align}
E(\{a_{\xi}\})=\sum_{n}  \left(f_n^{(\rho)} - \sum_{\xi} a_{\xi} f_{n}^{(\xi)} \right)^{\!2}\!,
\label{eq:functional}
\end{align}
subject to constraints $\sum_{\xi} a_{\xi} =1$ and $\sum_{\xi} a_{\xi} {\sigma}_{\xi} \geq 0$, which ensure the reconstructed operator corresponds to a physical state. This method only assumes that the unknown state lies within the calibration field of view defined by the probe states and can be  well-represented as a weighted sum of the probes states.

The physical origin for the FDP state estimation approach, encapsulated in the functional of Eq.~(\ref{eq:functional}),  can be understood by examining the Born rule, Eq.~(\ref{eq:bornrule}), applied to the probe states and estimated state. Minimization of the functional in Eq.~(\ref{eq:functional}) aims to reduce the difference between the probability distribution for the unknown state, $p_n^{(\rho)} \approx f_n^{(\rho)}$, and that calculated from the estimated state $p_n^{(\rho_{\rm{fdp}})} = \text{Tr}(\pi_n \rho_{\rm{fdp}}) = \sum_{\xi} a_{\xi} \text{Tr}(\pi_n \sigma_{\xi}) \approx \sum_{\xi} a_{\xi} f_n^{(\xi)}$, where probabilities are approximated by their corresponding measured frequencies.

To experimentally demonstrate the data-pattern calibration and its subsequent application to state reconstruction for a detector with a large number of outcomes we examine a balanced homodyne detector (BHD). This vital resource for continuous-variable (CV) quantum technologies is typically assumed to have an output that is proportional to the electric field quadrature of a well-defined spatial-temporal optical field mode \cite{LeonhardtBook,Lvovsky:09}. Indeed, balanced homodyne detection was used in the first quantum state and CV process estimation experiments \cite{Raymer:93,Lobino:08}. Optical interference of a reference beam called the local oscillator (LO) with the unknown signal field to be examined defines the detection mode, as presented in Fig.~\ref{fig:schematic}(b). The balanced scheme suppresses the LO technical noise and shot noise allowing ultra-sensitive sampling of the signal.  

The BHD is characterized by a set of probe states comprising 48 phase-averaged coherent states (PHAVs) with varying amplitudes. The probe states are generated deterministically by attenuating and phase-randomizing a laser beam, as shown in Fig.~\ref{fig:schematic}(c). The use of phase-averaged probes to characterize the detector implies that the calibration field of view enables access to the photon-number statistics of the unknown quantum states, which in the case of phase-invariant states constitutes complete quantum state estimation. 
 
For each probe state $\sigma_{\xi}$, an ensemble of $K=10^6$ optical pulses is measured sequentially by the BHD, yielding a set of voltages $\{V_1^{(\xi)},V_2^{(\xi)},...,V_K^{(\xi)}\}$ recorded by a fast oscilloscope. In the experiment the probe state amplitudes $|\alpha_{\xi}|$ range from $0.17$ to $2.24$ in approximately even steps of $0.043$. The frequency distribution $f_n^{(\xi)}$ of measurement outcomes $n$ for state $\sigma_{\xi}$ is formed by binning the voltage samples $\{V'\}$ after rescaling (see Methods, ``Balanced homodyne measurement''). 
Each frequency distribution consists of $151$ bins, implying as many detector outcomes, which is more than an order of magnitude greater than any QDT experiment to date. The procedure is repeated for each probe state $\sigma_{\xi}$ to give the detector data pattern set $\{f_n^{(\xi)}\}$, as shown in Fig.~\ref{fig:probes}(a), constituting the detector calibration. Note that no physical interpretation about the nature of the measurement outcomes is necessary with the FDP method \cite{Rehacek:10, Mogilevtsev:13}.

To demonstrate the FDP method for QSE, several non-classical photon-number (Fock) states \cite{Lvovsky:01, Zavatta:04, Ourjoumtsev:06, Cooper:13} are examined using the calibrated detector. The density matrices $\rho$ are estimated by fitting the probe state data patterns $\{f_n^{(\xi)}\}$ to the frequency distribution of measurement outcomes $f_n^{(\rho)}$ for heralded one-, two-, and three-photon Fock states according to Eq.~(\ref{eq:functional}). The Fock states are generated by pulsed spontaneous parametric down-conversion (SPDC) (see Methods, ``Fock state preparation'') and measured by the BHD. Frequency distributions $f_n^{(\rho)}$ for each Fock state are obtained in the same manner as for the probe states, shown in Fig.~\ref{fig:probes}. The functional defined in Eq.~(\ref{eq:functional}) is minimized for each state separately to determine the optimal set of coefficients $\{a_{\xi}\}$, shown in Fig.~\ref{fig:probes}(b-d).

The reconstructed Wigner functions $W(X,P)$ and photon number statistics $P(n)=\langle n | \rho^{\text{fdp}} | n \rangle$ for the generated Fock states are shown in Fig.~\ref{fig:density_matrices}. To compare with the commonly used maximum-likelihood (ML) quantum state estimation, the same homodyne data that yield the frequency distributions $f_n^{(\rho)}$, are used in a ML QSE algorithm \cite{Lvovsky:09}. Here a POVM of the form $\{\pi_{X}\}=\{|X\rangle\langle X|\}$ and phase-invariant states are assumed, where $|X\rangle$ is the quadrature eigenstate with eigenvalue $X$ (see Methods, ``Balanced homodyne measurement''). The photon number distributions reconstructed with the ML estimation are shown in Fig.~\ref{fig:density_matrices}. The fidelity $F=\rm{Tr}^2\left(\sqrt{\sqrt{\rho^{\text{fdp}}}\rho^{\text{ml}}\sqrt{\rho^{\text{fdp}}}}\right)$ between density matrices $\rho^{\text{fdp}}$ and $\rho^{\text{ml}}$  estimated by the FDP and ML methods for each Fock state is calculated, yielding fidelities $F_1=0.96\pm0.02$, $F_2=0.98\pm0.02$ and $F_3=0.92\pm0.02$ for the one-, two- and three-photon states respectively. This agreement between FDP and ML approaches indicates that the assumed quadrature POVM used in the ML estimation is accurate within the field of view experimentally examined by the probe state calibration set and experimental uncertainties. It also provides a self-consistency check between the two methods, thus demonstrating the applicability of FDP method to characterization of complex detectors and QSE. 

The ability to accurately calibrate the response of detectors is essential to the progress of quantum technologies. As quantum devices grow in size and complexity, so too will the measurement devices required for their operation and diagnosis. This necessitates the development of techniques to characterize increasingly elaborate detector responses. Although quantum detector tomography offers a possible route to achieve this goal, there are significant challenges to be met to accommodate detectors with increasing complexity. Here we have presented the first experimental demonstration of an alternative method for quantum detector characterization and its subsequent use in quantum state estimation by the fitting of data patterns. This approach enables calibration of complex, many-outcome detectors through direct, local measurement of the characteristic detector response to a set of probe states. The experimental calibration of a balanced homodyne detector and its subsequent use in reconstruction of non-classical Fock states presented here not only demonstrates successful implementation of a new detector characterization approach, but also goes an order of magnitude beyond any quantum detector characterization previously demonstrated. The data patterns approach is easily adapted to a variety of measurement devices and the experimental implementation presented shows its viability for complex detectors. We anticipate that this approach to detector calibration will become a standard approach to characterize measurement response in a local region of input space, adding to and complementing the global perspective of QDT.

\section*{Methods}
\textbf{Balanced homodyne measurement:} A time-domain balanced homodyne detector (BHD) with a bandwidth of $80$~MHz and signal-to-noise ratio of $14.5$~dB is used to perform measurements of the probe and signal fields \cite{Cooper_jmo:13}. For each incident optical pulse the BHD generates a voltage pulse. The BHD voltage $V$ is digitized by a computer-controlled digital storage oscilloscope (DSO). Drifts in the detector balance and gain are compensated by rescaling the measured voltage according to $V' = AV - B$. Rescaling parameters $A$ and $B$ are determined by acquiring pulses when the BHD input is blocked. Voltage samples with the input blocked are constrained to satisfy $\langle V' \rangle=C_1$ and ${\rm{var}}(V') = \langle V'^2 \rangle - \langle V' \rangle^2=C_2$, where $C_1$ and $C_2$ are constants, leading to the relations $A = \sqrt{(C_2 / {\rm{var}}(V))}$ and $B = A\langle V \rangle - C_1$. The LO repetition rate is $80$~MHz whereas the probe state repetition rate is approximately $4$~MHz. This enables continuous compensation for any drift \cite{Cooper_jmo:13}. In the case of the maximum-likelihood state reconstruction, where a POVM is assumed, $\langle V' \rangle=0$ and ${\rm{var}}(V')=1/2$ is used, making $V'$ correspond to a quadrature eigenvalue $X$ for the case of a perfect BHD. 

\textbf{Probe state preparation:} The phase-averaged coherent state probes are derived from a mode-locked Ti:Sapphire oscillator operating at a central wavelength of 830 nm, full-width half-maximum (FWHM) bandwidth of 10 nm, and repetition rate of $80$~MHz, Fig.~\ref{fig:schematic}(c). The beam is initially spatially filtered with a single-mode fiber (SMF) and spectrally filtered with an interference filter (IF) (Semrock LL01-830-12.5), to match the spatial-spectral mode of local oscillator. A computer-controlled motorized half-wave plate (MHWP) followed by a Glan-Taylor polarizer (GT) and calibrated neutral density (ND) filters enables precise control of the probe state amplitude. Phase averaging is achieved by driving a piezo-electric translator (PZT) on which one of the interferometer mirrors is mounted. The effective probe state amplitude $|\alpha_{\xi}|$ registered by the BHD is given by
\begin{align}
|\alpha_{\xi}|=\sqrt{P_{\text{meas}}\left(\frac{T}{R}\right) 10^{-\text{OD}_1-\text{OD}_2} \mathcal{V}^2 \frac{\lambda_0}{hc\nu} },
\label{eq:alpha}
\end{align}
where $P_{\text{meas}}$ is the average power measured on the calibrated power meter (PM) (NIST-traceable Coherent FieldMaxII-TO power meter), $R$($T$) is the reflectivity (transmissivity) of the beam splitter (BS), $\text{OD}_{1(2)}$ is the optical density of filter 1(2), $\nu$ is the laser repetition rate, $\lambda_0$ is the laser central wavelength and $\mathcal{V}$ is the interference visibility between the LO mode and probe state mode. The interference visibility is measured by removing the ND filters and directing one output of the 50:50 beam splitter used for homodyne detection using a flip-mirror to a fast photodiode that records the classical interference pattern as the phase is modulated. A calibrated Thorlabs CCS175 spectrometer (CS) measures the central wavelength $\lambda_0$. The probe state spectrum has a central value $\lambda_0 = 827.6$~nm. The laser repetition rate $\nu$ is measured using a high-precision frequency counter. The repetition rate is decreased by a pulse-picker and was measured to be $3.997$~MHz for the duration of the probe state measurement.

\textbf{Fock state preparation:} A two-mode squeezed vacuum (TMSV) state of the form $|\psi\rangle\approx\sqrt{1-\gamma^2}\left(|0,0\rangle + \gamma |1,1\rangle + \gamma^2 |2,2\rangle + \gamma^3 |3,3\rangle + O(\gamma^4)\right)$, where $\gamma$ is the squeezing parameter and $|n,m\rangle$ is a two-mode state with $n$ ($m$) photons in the signal (trigger) mode, is generated by type-II pulsed spontaneous parametric down-conversion (SPDC) in a bulk potassium di-hydrogen phosphate (KDP) crystal. The signal and trigger modes are separated by a polarizing beam splitter (PBS). A spatially multiplexed detector (SMD) comprising a fiber beam splitter network and three avalanche photodiodes (APDs) enables heralding of one-, two- and three-photon Fock states conditioned on registering one, two or three ``clicks" respectively from the SMD \cite{Cooper:13}. The heralded Fock state in the signal mode is combined with the LO for performing balanced homodyne detection.

\textbf{Reconstruction and uncertainty estimation:} The uncertainties in the state fidelities between ML and FDP approach are estimated by taking into account both statistical and systematic sources of error. Statistical errors arise due to the finite number of measurement events used to generate data patterns and are given by $\sqrt{N_n}$ for each bin $n$ with population $N_n$ in the histograms forming measurement outcome frequency distributions $f_{n}^{(\xi)}$ and $f_n^{(\rho)}$. The bin width is chosen such that each bin is sufficiently narrow to adequately sample the underlying probability distribution \cite{Rehacek:10}. For the FDP state reconstructions, there are also errors originating from the estimates of $P_{\text{meas}}$ and $\mathcal{V}$ used to determine $|\alpha_{\xi}|$ for each probe state. Uncertainties of 5\% in $P_{\text{meas}}$ and 1\% in $\mathcal{V}$ are estimated, which gives a total error for each $|\alpha_{\xi}|$ of $2.7$\%. For the ML state reconstructions, the primary source of error is the uncertainty in detector efficiency $\eta_{\text{bhd}}$. This must be independently determined and explicitly taken into account in the ML estimation to compare with the FDP method, which automatically incorporates the detector efficiency. Monte Carlo simulation enables estimation of the uncertainties for the estimated state density matrix elements and hence also the fidelities when comparing the FDP approach with the results from ML reconstruction. The Monte Carlo simulation takes into account statistical errors in the data patterns and state frequency distributions, as well as the uncertainty in the probe state amplitudes $|\alpha_{\xi}|$ which impacts the probe state density matrices $\{\sigma_{\xi}\}$ and hence the estimated state $\rho^{\text{fdp}}$, Eq.~(\ref{eq:estimatedstate}).

\section*{Author contributions}
M.~C.\ and B.~J.~S.\ conceived the project. M.~C.\ designed and performed the experiment. M.~C.\ and M.~K.\ performed modelling and data analysis including code to implement the algorithm. All authors contributed equally to writing the manuscript.

\section*{Acknowledgements}
We are grateful for helpful discussions with M.~G.\ Raymer and M.\ Barbieri. This work was supported by the University of Oxford John Fell Fund and EPSRC grant No.\ EP/E036066/1. M.~K.\ was supported by a Marie Curie Intra-European Fellowship 301032 (CV-QDAPT) within the European Community 7th Framework Programme.


\end{document}